\documentstyle[graphicx,12pt]{mrsproc}

\font\timesrm=ptmr at 12pt
\font\timesit=ptmri at 12pt
\font\timesbf=ptmb at 12pt

\renewcommand{\normalsize}{\fontsize{12pt}{12pt}\selectfont}
\normalsize

\renewcommand{\rm}{\timesrm}\rm
\renewcommand{\it}{\timesit}
\renewcommand{\bf}{\timesbf}

\begin{document}

\begin{center}
{\bf Quasicrystal approximants with novel compositions and structures}
\end{center}
\bigskip

\raggedright

M. Mihalkovi\v{c}$^1$ and M. Widom,\\
Department of Physics, Carnegie Mellon University, Pittsburgh, PA  15213\\
$^1$also at: Institute of Physics, Slovak Academy of Sciences, 84228 Bratislava, Slovakia

\parindent0.3in

\begin{abstract}
We identify several new quasicrystal approximants in alloy systems in
which quasicrystals have not been previously reported.  Some occur in
alloys with large size contrast between the constituent elements,
either containing small Boron atoms, or large Ca/Eu atoms, leading to
quasicrystal structures quite different from currently known systems
where the size contrast is smaller.  Another group of the approximants
are layered Frank--Kasper structures, demonstrating competition
between decagonal and dodecagonal ordering within this family of
structures.
\end{abstract}

\section{INTRODUCTION}
Axial quasicrystals are structures that possess one axis of
crystallographically forbidden rotational symmetry.  The structure
is quasiperiodic in the plane perpendicular to this axis, but can
be periodic in the direction parallel to the axis.  Since they never
exactly repeat, their lattice constant can be thought of as infinite
within the quasiperiodic plane.  Along the periodic direction the
lattice constant is finite and might be fairly small.

Quasicrystal approximants are ordinary, though complex, crystals whose
local structural motifs naturally extend to quasiperiodic structures
with the crystallographically forbidden rotational
symmetries~\cite{QCApprox}.  Approximants reproduce within their unit
cell a portion of an aperiodic structure, certain special fragments of
which can be extended periodically while maintaining reasonable local
atomic structures.  Generally approximants have large lattice
parameters in the directions in which the corresponding quasicrystal
is quasiperiodic.
  
Structures that possess at least one large lattice parameter are
thus candidates for being a quasicrystal approximant.  If, in
addition they possess a short lattice parameter orthogonal to their
large one(s) this direction is a candidate for being an
axis of high rotational symmetry in an axial quasicrystal.

Experimentally observed axial quasicrystal symmetries are octagonal
(8x), decagonal (10x) and dodecagonal (12x).  To-date stable decagonal
quasicrystals are known in the compounds AlCoNi, AlCoCu, AlMnPd,
AlCrNi, AlNiRu, AlCuRh, AlFeNi, ZnMgDy, AlCrFe, GaFeCuSi, AlCuCr and
possibly others.  In all these cases, the structures are believed to
be layered, with a basic 4~\AA~ unit consisting of two atomic layers
that are either stacked periodically or with some modulation leading
to periodicities of 4, 8, 12 and 16~\AA~.  Perpendicular to the axis
the structures are quasiperiodic.  Some observed metastable dodecagonal
quasicrystals occur in Ta-Te, Ni-Cr, Bi-Mn, Ni-V.

In the process of searching for bulk metallic glass-forming
compounds~\cite{alcacu_unp,bfezry_unp} we developed a database of
intermetallic structures which now contains about 1000 structures
drawn mainly from standard
references~\cite{TernaryPD,Pearson,BinaryPD,PD_new}.  The data base
contains the Pearson symbol of each structure (e.g. B$_4$Mg$_2$Ru$_5$
has Pearson symbol oP22 indicating an orthorhombic Primitive cell with
22 atoms), the space group (e.g. number 55, or Pbam), as well as the
lattice parameters and Wyckoff coordinates.

We screened this database for structures that contain one short
lattice parameter of 5~\AA~ or less and one long lattice parameter of
8~\AA~ or more.  Owing to the presence of at least one long axis, these
structures tend to have relatively large numbers of atoms per unit
cell, as quoted in their Pearson symbols. We drew pictures of the
structures projected along their short axis, then visually examined
the picture.  In many cases we observed the presence of local
structural motifs of approximate 5-fold symmetry.  We took special
note of compounds that were chemically similar to the glass-forming
compounds of interest (Al-Ca-Cu and B-Fe-Y-Zr).

When we found local 5-fold symmetry we attempted to formulate a tile
decoration model that would reproduce the observed structure and allow
us to extend the structure to larger approximants or a complete
quasicrystalline structure.  We were able to do this in several cases.

Given a proposed quasicrystal model, we carried out total energy
calculations using PAW potentials~\cite{PAW} in the generalized gradient
approximation~\cite{PW91} as implemented by the program
VASP~\cite{VASP,VASP2}.  Total energies of compounds of differing
composition can be compared by measuring all energies relative to the
tie-line connecting cohesive energies of pure elements.  All
structural energies were for fully relaxed structures and were
converged to a precision of 1 meV/atom.  Because we relaxed the atomic
volume, these energy differences are the structural enthalpy of
formation, $\Delta H_{for}$, evaluated at temperature T=0K.

\section{BORON-CONTAINING PHASES}
\begin{figure}
\includegraphics[width=3 in]{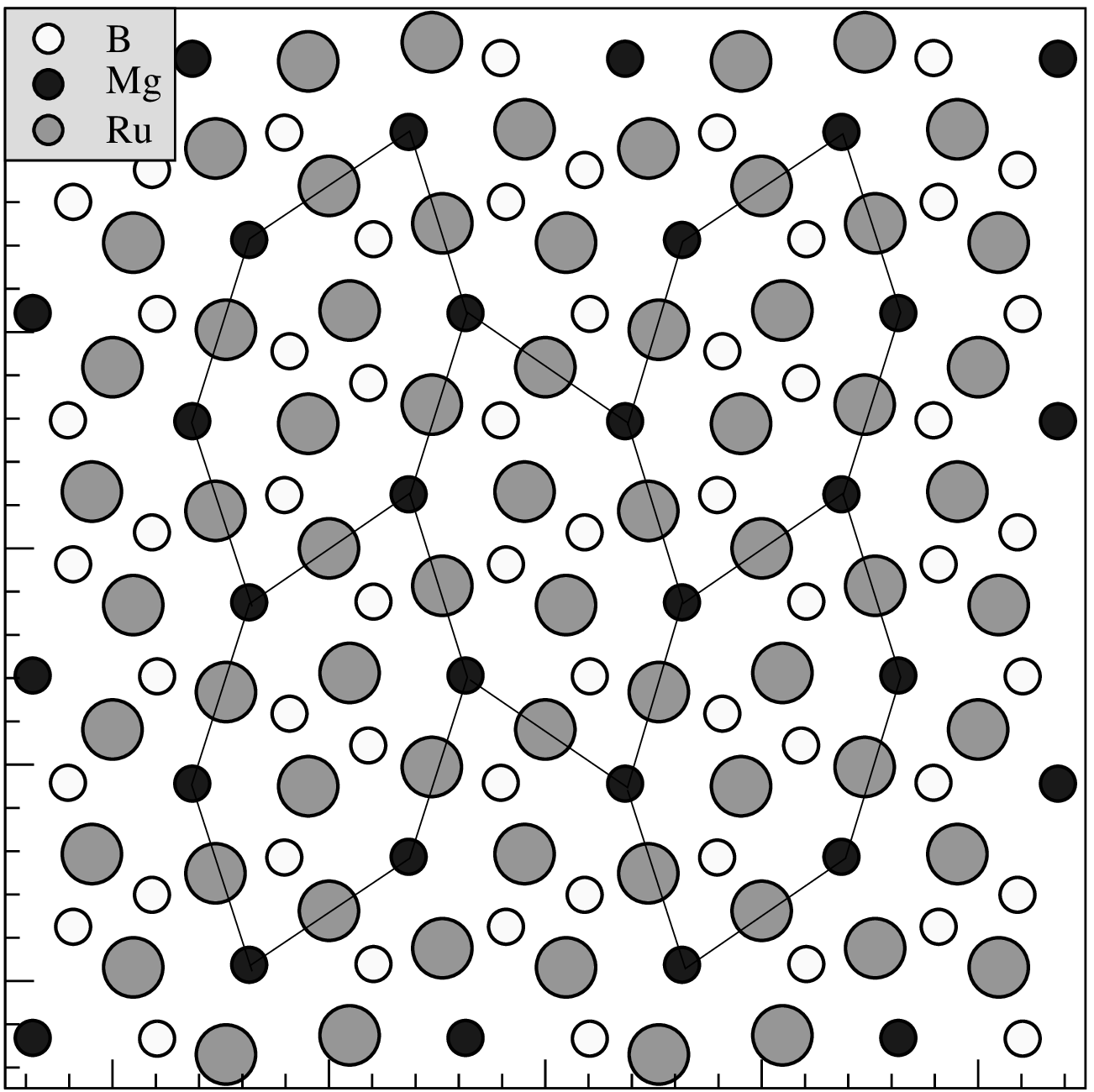}
\includegraphics[width=3 in]{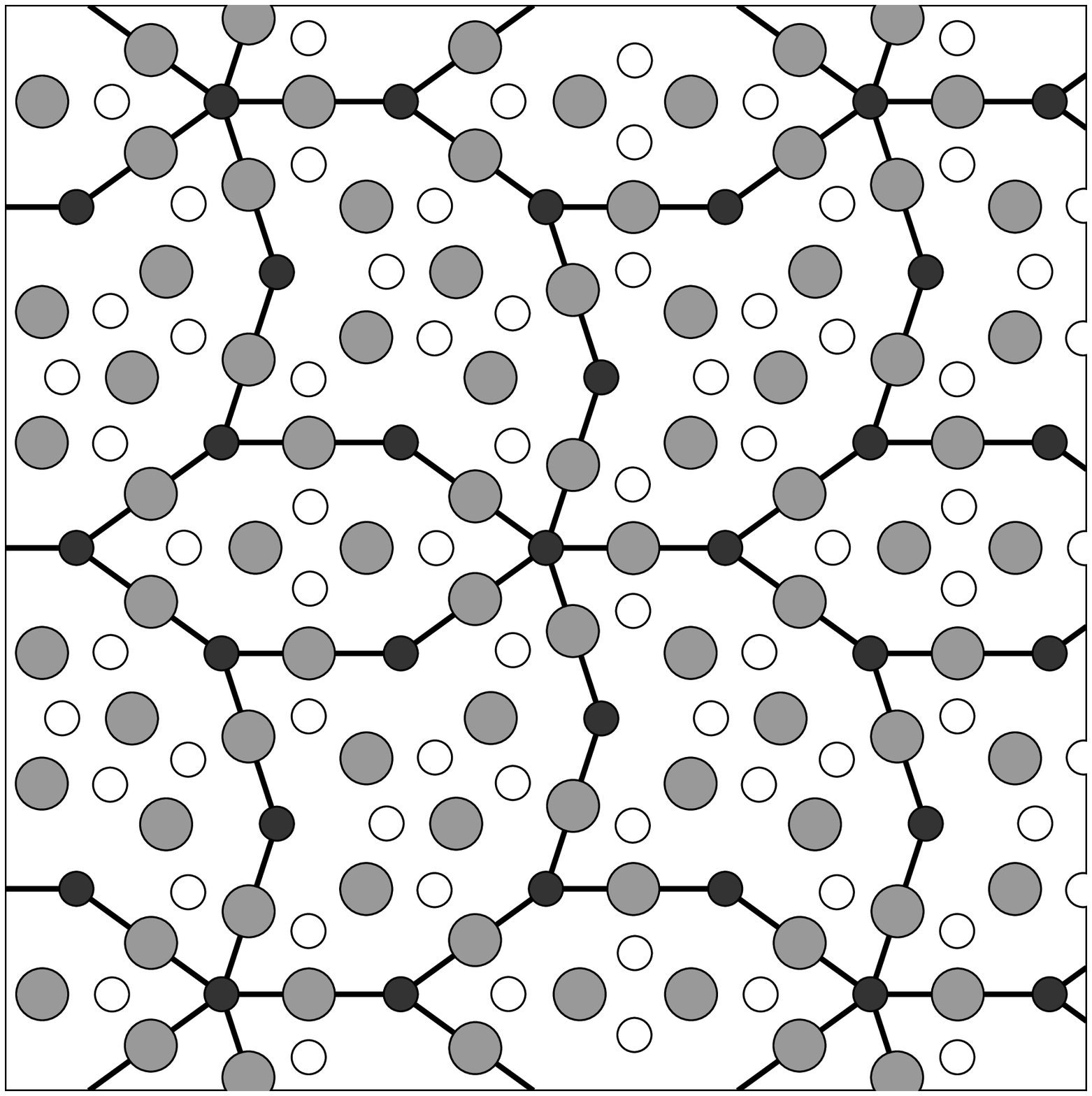}
\caption{\label{fig:bmgru}(a)B$_4$Mg$_2$Ru$_5$ approximant. (b)
Proposed model quasicrystal. Large and small circles indicate upper
and lower layers.  Black, gray and white indicate large, medium and
small atoms, respectively.}
\end{figure}

One interesting case is B-Mg-Ru alloys, where B$_4$Mg$_2$Ru$_5$.oP22
and B$_{11}$Mg$_5$Ru$_{13}$.oP62 are two observed
structures~\cite{oP22oP62} that we believe are approximants of a
B$_{38}$Mg$_{17}$Ru$_{45}$ decagonal quasicrystal.
Figure~\ref{fig:bmgru} shows the structure of oP22 and our proposed
extension to a decagonal quasicrystal based on boat-hexagon tilings.

This compound is chemically unlike any presently known for
quasicrystal formation.  Notably, it contains a fairly large
concentration of Boron.  Owing to the prevalence of icosahedra in
elemental Boron, researchers have long suggested that icosahedral
quasicrystals might occur in Boron-rich
compounds~\cite{Favio,Takeda,Weygand,Kimura,ZH,Boustani}, but none has
yet been found.

Our proposed structure is promising because we have investigated the
enthalpy of formation of other, larger hypothetical approximants.  We
consider how far these approximants lie from the convex hull of
enthalpy versus composition for all known B-Mg-Ru structures, and find
a large set of structures, nearly degenerate in energy, lying slightly
above the convex hull.  This is just the scenario envisioned for the
entropic stabilization of quasicrystals, so we suggest that perhaps
this compound will exhibit a thermodynamically stable high-temperature
decagonal quasicrystal phase.  We will describe the tiling energetics,
including an evaluation of the energy of vertical
structural degrees of freedom, breaking rigid 2-layer periodicity
along the stacking (shortest periodic) axis.
in an upcoming publication~\cite{MW_unp}.

Likewise, Al$_8$CaCo$_2$.oP44 is an approximant of a hypothetical
Al$_x$Ca$_y$Co$_z$ decagonal phase.  The structure here is quite
similar to B-Mg-Ru, with the substitutions Mg$\rightarrow$Ca,
B$\rightarrow$Co exactly reproducing the B-Mg layer as a Co-Ca layer.
However, the Al-content in the adjoining atomic layer is about
3$\times$ the Ru content of the adjoining layer in the B-based
compound.  Unlike the case of B-Mg-Ru, where the star (S) tile is
energetically unfavorable, leading to tilings of hexagons and boats
only, for Al-Ca-Co the star tile is comparable in energy to the boat
(B) and hexagon (H) tiles, leading to full Hexagon-Boat-Star tilings.


Due to the interest in finding genuinely B-rich quasicrystals, we
screened our database to select high B concentration and found several
candidate approximants.  Noteworthy cases include B$_6$ReY$_2$.oP36,
B$_7$ReY$_3$.oC44 and B$_4$CrY.oP24.  All of these feature planar
Boron pentagons, centered by a medium-sized transition metal.  In the
first two cases, a second pentagonal ring of Y-atoms surrounds the
pentagonal ring of B-atoms.

We have not yet identified the proper manner to extend these
structures to a full decagonal quasicrystal in a manner that achieves
sufficiently low energy, but since they are all stable as BFeY alloys,
and occur close-by in composition, we suspect there may be a unifying
quasicrystal model to which they are approximants.

The most promising of these cases is B$_7$ReY$_3$, whose structure may
be interpreted in terms of a pure tiling of skinny rhombuses, with B-Y
pentagonal clusters centered by Re decorating tiling vertices, and
alternating heights along short periodic direction along the tiling
edges.  Our best attempt extending this decoration to the fat tile is
unstable by a moderate 20meV/atom energy.  B$_2$C$_2$Sc.oP20, another
curious structure, is isostructural with B$_4$CrY.oP24 except that the
pentagonal rings contain no medium-sized transition metal. In this
compound, C and B share the pentagonal rings.

\section{RARE-EARTH-CONTAINING PHASES}
\begin{figure}
\includegraphics[width=3 in]{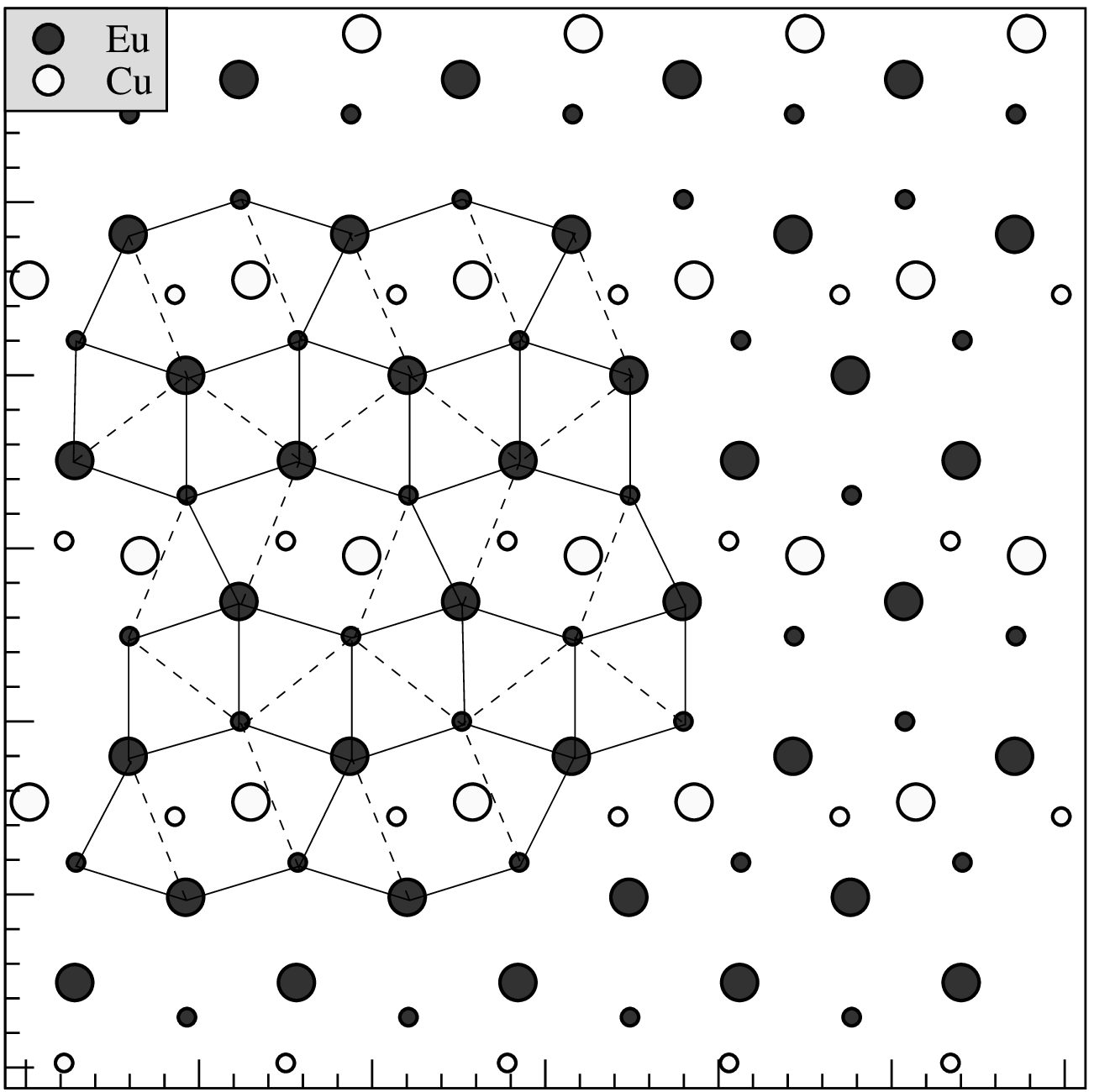}
\includegraphics[width=3 in]{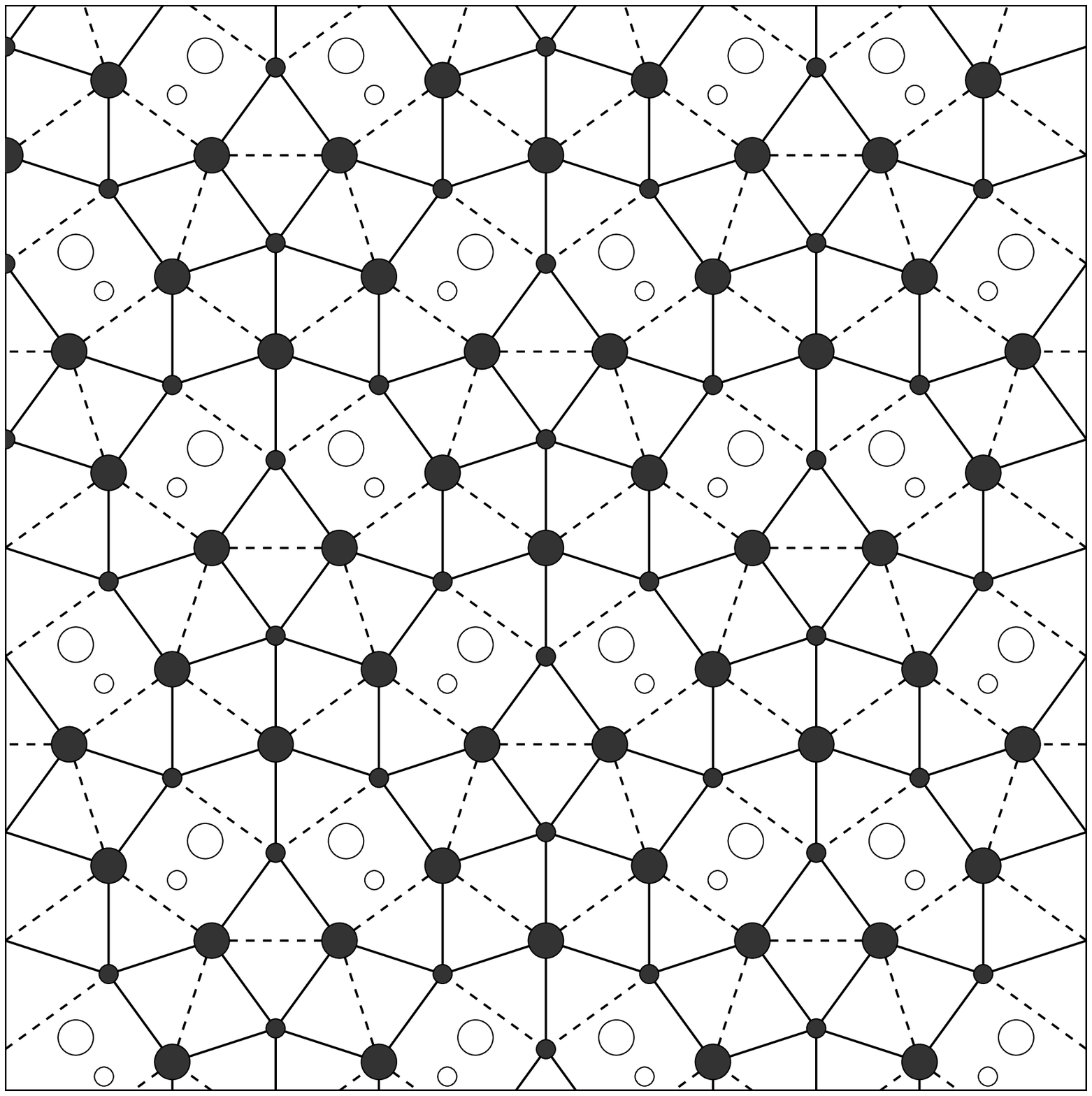}
\caption{\label{fig:cueu}(a) CuEu$_2$ approximant. (b) Cu$_4$Eu$_9$
approximant that we predict to be stable}
\end{figure}

Another case that we consider promising for the occurence of an
entropically stabilized decagonal quasicrystal is Eu-Cu and the
isostructural compound Ca-Cu.  Here again we find two experimentally
observed approximants, CuEu.oP8
and CuEu2.oP12 that allow us to
devise decoration rules and extend the motif to a full decagonal
quasicrystal, as shown in Fig.~\ref{fig:cueu}.  Surprisingly, one of
our larger approximant structures that we produced by hand,
Cu$_4$Eu$_9$, turns out also to be predicted stable, though it has not
been reported experimentally.  The remaining approximants we tested
lie slightly above the convex hull.  In order for the quasicrystal to
be stabilized entropically a mechanism introducing structural degrees
of freedom along vertical direction is needed.

As in the case of the B-Mg-Ru system, this new alloy system is
chemicaly unlike previously known quasicrystals, this time because it
is rich in a Rare Earth element. An unique feature of this potential
quasicrystal is striking simplicity of the decoration, with tiling
edge length equal to the nearest neighbor interatomic distance.  The
small tile size is responsible for the large composition range of the
family of approximants, spanning $x_{Eu}$=0.5$\rightarrow$1 range of
the Eu/Ca content (interestingly, both orthorhombic and monoclinic
CaCu structures, as well as CuEu with BFe structures, are valid
approximants of the same quasicrystal family).

Unlike the Boron-based structures, where the basic tiles were drawn from
HBS tilings, for Cu-Eu the basic structure is based on a decoration of
the rectangle-triangle (RT) tiling.  The rectangle-triangle tiling is
a 10$\times$ symmetric variant of the 12$\times$ symmetric
square-triangle tiling which form dodecagonal quasicrystals.
Interestingly, both these models have exact solutions for their
entropy and phason elastic constants~\cite{exact}.  They also share
the property that their phason degrees of freedom in general are quite
complex involving non-local ``zipper'' updates~\cite{zipper}.

\section{FRANK-KASPER STRUCTURES}

The final class of structures we mention is the Frank-Kasper family,
characterized by tetrahedral close-packing.  
Constraints of close packing dictate that only special atomic
environments with triangulated coordination shells occur, namely
$Z$=\{12,14,15,16\} with 12-16 nearest neighbors surrounding the central atom. 
Icosahedral quasicrystal
Frank-Kasper phases are represented by prototype
Al$_{6}$Mg$_{11}$Zn$_{11}$.cI162 phase~\cite{HE86}, while the
Mg$_4$Zn$_7$.mC110 phase is an approximant of a decagonal quasicrystal.
The Laves phase of MgZn$_2$.hP12 may be viewed as a common approximant
for both quasicrystal Frank--Kasper families.

The quasicrystal Frank-Kasper structures are pseudobinary, in that large atoms
(e.g. Mg) center highly-coordinated polyhedra ($Z$=14, 15 or 16) while
smaller atoms (e.g. Al or Zn) have icosahedral $Z$=12 coordination.
In both icosahedral and decagonal structures about 60\% of the atoms
are of the smaller $Z$=12 coordinated type. The close relationship
between decagonal and icosahedral FK structures was discussed by Roth
and Henley~\cite{RH97}.


\begin{figure}
\includegraphics[width=3 in]{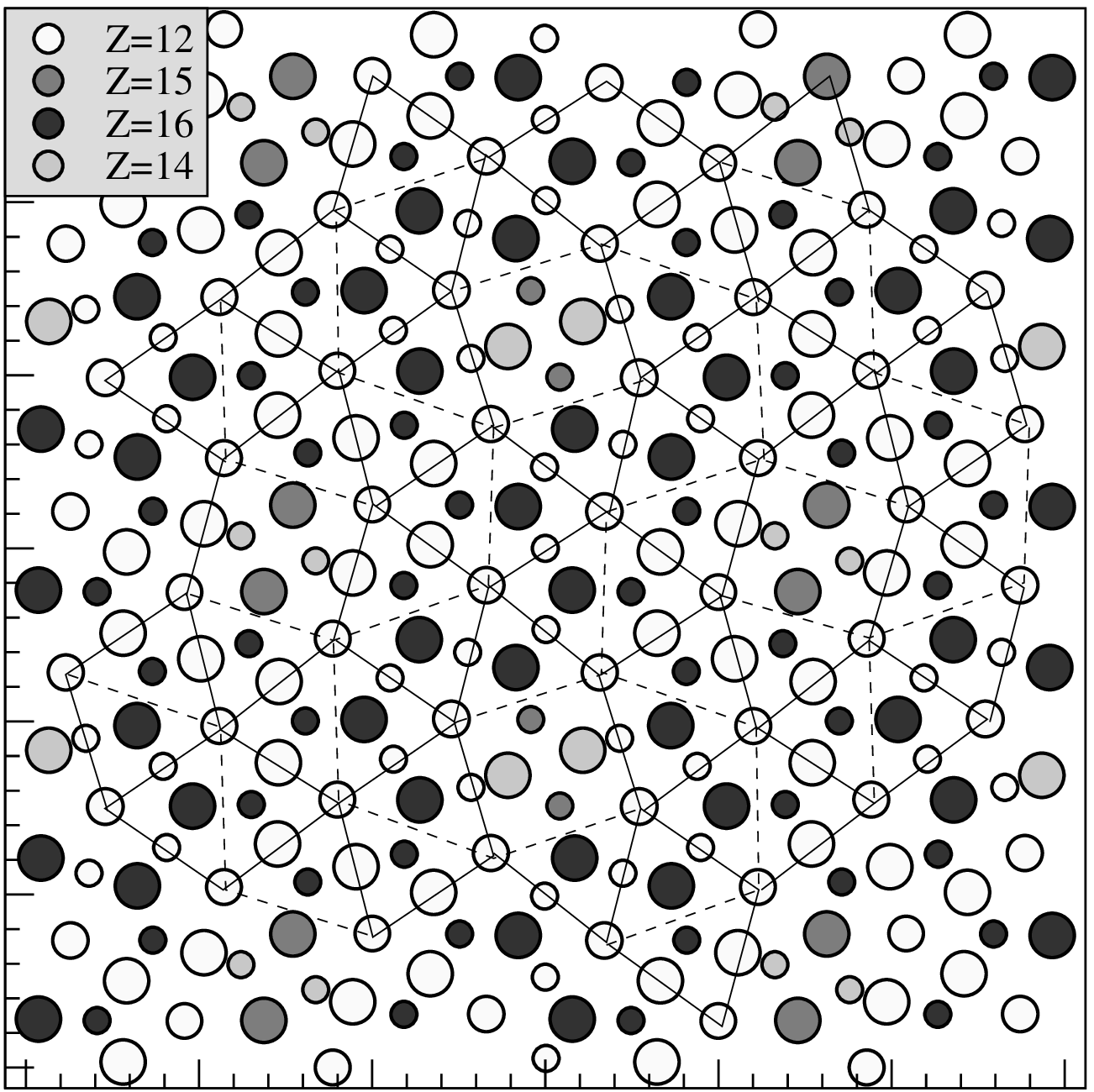}
\includegraphics[width=3 in]{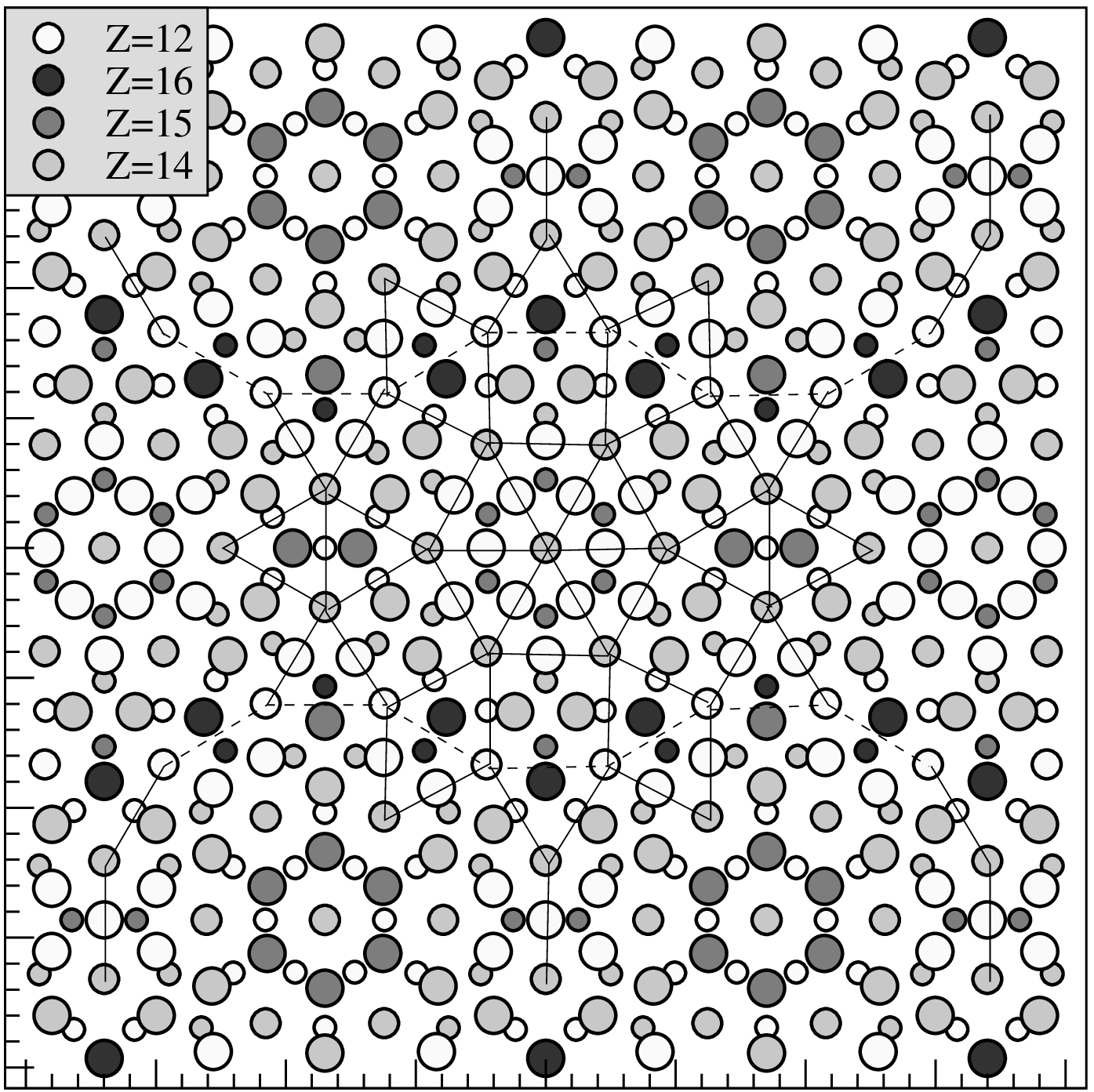}
\caption{\label{fig:fk}(a) Decagonal Co$_8$Mn$_9$Si$_3$  and (b) dodecagonal
Mn$_4$Si structures. Tile edges: $e_1$ (solid) $e_2$ (dashed).}
\end{figure}

We identified a family of layered (short axis $c\sim$5\AA)
close-packed structures that interpolates between decagonal and
dodecagonal ordering. In both cases, the structures may be viewed as
simple decorations of square-triangle tiling (for the moment, we don't
distiguish between isosceles/equilateral (60$^{\circ}$/72$^\circ$)
triangles and between squares/rectangles).  Tiling vertices are
occupied by pairs of atoms at $z=\pm$1/4.  Two kinds of tiling edges
connect nearest-neighbor vertices: $e_1$ with an atom decorating the
bond midpoint; $e_2$ containing no atom.  Each triangle is decorated
by one interior atom, each rectangle by four.

In the limit of pure dodecagonal order, {\em all} edges are of type
$e_1$.  Since all edges are equivalent, the triangles become
equilateral and the rectangles are actually square.  In the opposite
limit of pure decagonal order, each triangle has two $e_1$ and one
$e_2$ edge (so they are isosceles), and each rectangle has two $e_1$
and two $e_2$ edges.  Squares possess a 4$_2$ screw axis that
rectangles lack.  A packing rule requires adjacent tiles to share
their common edge type.

Conversion of $e_2$ edges into $e_1$ by placing extra atoms on their
midpoints converts $Z$=12 vertex polyhedra into $Z$=14.  We
define a parameter to measure the degree of dodecagonal {\em vs.}
decagonal order in a layered FK approximant.  First, identify the tile
vertices by locating atom pairs at $z=\pm$1/4 along the short
axis. They must have coordination polyhedron $Z$=12 or $Z$=14. Then,
define
\begin{equation}
  \zeta = N_{14}/(N_{12}+N_{14})   \label{eq:zeta}
\end{equation}
where $N_{Z}$ denote numbers of tiling vertices with $Z$=12 or $Z$=14
coordination. A structure is pure dodecagonal for $\zeta$=1, and pure
decagonal for $\zeta$=0.  Conversion of $e_2\rightarrow e_1$ produces
$c$-height alternation conflicts for neighboring atoms inside the
tiles, so it must be done non-locally.

Table~\ref{tab:fk} summarizes our results for several structures.
Besides the $\zeta$ parameter, we provide numbers of triangles per
unit cell $N_t$ and the $N_t$/$N_s$ ratio, where $N_s$ is number of
``squares''.  The ideal $N_t$/$N_s$ ratios for decagonal and
dodecagonal quasicrystal are 2($\sqrt{5}+1$)$\sim$6.47 and
4/$\sqrt{3}\sim$2.31 respectively.  Structures with $\zeta>0$ have
$N_t/N_s$ close to 2, while the ratio is much larger than 2 for all
but the smallest structures with $\zeta=0$.

\begin{table}
\caption{\label{tab:fk} Dodecagonal and decagonal 
Frank-Kasper approximants}
\begin{tabular}{l|c|c|c||l|c|c|c}
structure & $N_t$ & $N_t$/$N_s$ & $\zeta$ &
structure & $N_t$ & $N_t$/$N_s$ & $\zeta$ \\
\hline
Mg$_4$Zn$_7$.mC110               & 32 & 16     & 0  &  
Co$_{17}$Si$_{13}$V$_{20}$.mC50  & 12 &  6     & 0  \\  
Co$_8$Mn$_9$Si$_3$.oP74          & 20 & 10     & 0  &  
Al$_3$Nb$_{10}$Ni$_9$.oP52       &  8 &  2     & 0  \\  
Cr$_9$Mo$_{21}$Ni$_{20}$.oP56    &  8 &  2     & 0.5&  
Mn$_4$Si.oI186                   & 14 & 2.333  & 0.692\\ 
CrFe.tP30                        &  4 &  2     & 1  &&&&\\  
\end{tabular}
\end{table}

Fig.~\ref{fig:fk} shows as examples a decagonal structure
(Co$_8$Mn$_9$Si$_3$.oP74), and a dodecagonal
(Mn$_4$Si.oI186).  The $e_1$ edges are shown as solid lines, $e_2$
dashed.  While Mn$_4$Si exhibits some icosahedra at the
tiling vertices ($\zeta\sim$0.7), we regard it as a prominent
dodecagonal approximant, due to the presence of large 12-fold
clusters.  These clusters actually extend beyond the dodecahedra
outlined in the figure, so the clusters overlap and {\em cover} the
structure.

Finally, note that the decagonal limit is optimal for pseudo-binary
compounds with about $\sim$40\% large and 60\% small ($Z$=12)
atoms. In the compounds listed in Table~\ref{tab:fk} with $\zeta$=0,
the elements Mg, Nb, V and Mn play role of the larger atoms, and
occupy centers of $Z>12$ polyhedra.  As $\zeta\rightarrow$1, the
fraction of $Z$=12 polyhedra decreases to about 1/3, by converting
$Z$=12 to $Z$=14.  Meanwhile, volumes of the highly-coordinated
polyhedra are only marginally bigger than those of the icosahedra, so
that the dodecagonal structures are pseudo-monoatomic, by which we
mean the atomic size contrast is small.

\section{ACKNOWLEDGEMENTS}
We wish to acknowledge useful discussions with C.L. Henley.  This work
was supported by NSF grant DMR-0111198 and benefited from metallic
glass research funded by DARPA.


\begin{thebibliography}{0}
\bibitem{QCApprox} V. Elser and C.L. Henley, Phys. Rev. Lett.~{\bf 55}, 2883-6 (1985).
\bibitem{alcacu_unp} M. Gao, G. Shifflet, M. Mihalkovi\v{c} and M. Widom, unpublished, (2003). 
\bibitem{bfezry_unp} M. Mihalkovi\v{c} and M. Widom, in preparation, (2003).
\bibitem{TernaryPD} P. Villars, A. Prince and H. Okamoto, {\it Handbook of ternary alloy phase diagrams}, ASM International, Materials Park, Ohio, (1995).
\bibitem{Pearson} P. Villars, {\it Pearson's Handbook, Desk Edition}, ASM International, Materials Park, Ohio, (1997).
\bibitem{BinaryPD} {\it Binary Alloy Phase Diagrams}, edited by T.B. Massalski, et al., ASM International, Materials Park, Ohio, (1990).
\bibitem{PD_new} {\it Desk Handbook: Phase Diagrams for Binary Alloys}, edited by H. Okamoto, ASM International, Materials Park, Ohio, (2000).
\bibitem{PAW} G. Kresse and J. Joubert, Phys. Rev. B~{\bf 59}, 1758 (1999).
\bibitem{PW91} J.P. Perdew and Y. Wang, Phys. Rev. B~{\bf 45}, 13244 (1992).
\bibitem{VASP} G. Kresse and J. Hafner, Phys.\ Rev. B~{\bf 47}, RC558 (1993).
\bibitem{VASP2} G. Kresse and J. Furthmuller, Phys. Rev. B~{\bf 54}, 11169 (1996).
\bibitem{oP22oP62} K. Schweitzer and W. Jung, Z. Anorg. Allg. Chemie~{\bf 530}, 127-134 (1985).
\bibitem{Favio} P. Favio {\it et al}, Micros. Microan. Microstruct.~{\bf 7}, 225-34 (1996).
\bibitem{Takeda} M. Takeda {\it et al}, {\it 5$^{th}$ International Conference on Quasicrystals}, edited by edited by H. Okamoto, , World Scientific, Singapore, Materials Park, Ohio, (1995).
\bibitem{Weygand} C. Weygand and J.-L. Verger-Gaugry, C. R. Acad Sci. II~{\bf 320}, 253-7 (1995).
\bibitem{Kimura} K. Kimura, Mat. Sci. Eng. B~{\bf 19}, 67-71 (1993).
\bibitem{ZH} W.-J. Zhu and C. L. Henley, Europhys. Lett.~{\bf 51}, 133-9 (2000).
\bibitem{Boustani} I. Boustani, A. Quandt and P. Kramer, Europhys. Lett.~{\bf 36}, 583-8 (1996).
\bibitem{MW_unp} M. Mihalkovi\v{c} and M. Widom, Unpublished~{\bf },  (2003).
\bibitem{exact} B. Nienhuis, Phys. Rep.~{\bf 301}, 271-92 (1998).
\bibitem{zipper} M. Oxborrow and M. Mihalkovi\v{c}, {\it Aperiodic '97}, edited by M. de Boissieu, J.-L. Verger-Gaugry and R. Currat, World Scientific, Materials Park, Ohio, (1997).
\bibitem{HE86} C. L. Henley and V. Elser, Phil. Mag. B~{\bf 53}, L59 (1986).
\bibitem{RH97} J. Roth and C. L. Henley, Phil. Mag. A~{\bf 75}, 861 (1997).
\end{thebibliography}
\end{document}